\documentclass[sn-standardnature]{sn-jnl}

\jyear{2022}%

\theoremstyle{thmstyleone}%
\theoremstyle{thmstyletwo}%

\theoremstyle{thmstylethree}%

\usepackage{float} 
\usepackage{lipsum}
\usepackage{epstopdf} 
\usepackage{amsmath}
\usepackage{breqn}
\usepackage[section]{placeins}
\usepackage{xcolor} 

\begin{document}

\title[Petviashvili Method for the Fractional Schr\"{o}dinger Equation]{Petviashvili Method for the Fractional Schr\"{o}dinger Equation}


\author*[1,2,3]{\fnm{Cihan} \sur{Bay\i nd\i r}}\email{cbayindir@itu.edu.tr}

\author[2]{\fnm{Sofi} \sur{Farazande}}\email{sofi.farazande@boun.edu.tr}

\author[4]{\fnm{Azmi Ali} \sur{Altintas}}\email{altintas.azmiali@gmail.com}

\author[5,3]{\fnm{Fatih} \sur{Ozaydin}}\email{fatih@tiu.ac.jp}

\affil*[1]{\orgdiv{Engineering Faculty}, \orgname{\.{I}stanbul Technical University}, \orgaddress{\street{Maslak}, \city{\.{I}stanbul}, \postcode{34469}, \country{Turkey}}}

\affil[2]{\orgdiv{Engineering Faculty}, \orgname{Bo\u{g}azi\c{c}i University}, \orgaddress{\street{Bebek}, \city{\.{I}stanbul}, \postcode{34342},  \country{Turkey}}}

\affil[3]{\orgname{CERN}, \orgaddress{\street{1211}, \city{Geneva}, \postcode{23},  \country{Switzerland}}}

\affil[4]{\orgdiv{Department of Physics, Faculty of Science}, \orgname{\.{I}stanbul University}, \orgaddress{\street{Vezneciler}, \city{\.{I}stanbul}, \postcode{34116},  \country{Turkey}}}

\affil[5]{\orgdiv{Institute for International Strategy}, \orgname{Tokyo International University}, \orgaddress{\street{1-13-1 Matoba-kita, Kawagoe}, \city{Saitama}, \postcode{350-1197},  \country{Japan}}}


\abstract{In this paper, we extend the Petviashvili method (PM) to the fractional nonlinear Schr\"{o}dinger equation (fNLSE) for the construction and analysis of its soliton solutions. We also investigate the temporal dynamics and stabilities of the soliton solutions of the fNLSE by implementing a spectral method, in which the fractional-order spectral derivatives are computed using FFT routines, and the time integration is performed by a $4^{th}$ order Runge-Kutta time-stepping algorithm. We discuss the effects of the order of the fractional derivative, $\alpha$, on the properties, shapes, and temporal dynamics of the solitons solutions of the fNLSE. We also examine the interaction of those soliton solutions with zero, photorefractive and q-deformed Rosen-Morse potentials. We show that for all of these potentials the soliton solutions of the fNLSE exhibit a splitting and spreading behavior, yet their dynamics can be altered by the different forms of the potentials and noise considered. }




\maketitle

\section{Introduction}
Fractional calculus has been attracting an increasing attention due its vast applications.
In hydrodynamics, fractional mass conservation~\cite{wheatcraft2008fractional} uses the advantage of fractional Taylor series~\cite{li2021mass} in differentiation. 
Groundwater flow engineering is one of the most widely studied applications of fractional calculus. 
For instance, some analytical solutions of time-fractional Thiem and Theis groundwater flow equations are derived in~\cite{atangana2013use}, and fractional advection-dispersion models that are applicable to different soil types and conditions are studied in~\cite{lu2020quantifying}. 
For continuum studies, some general fractional constitutive mass flux density vectors are studied in~\cite{mirza2021analytical}, which employ standard techniques like Laplace, sine-Fourier, and cosine-Fourier transforms. 
In~\cite{mesgarani2021numerical}, fractional advection-dispersion models are studied numerically, that uses Caputo derivatives for the approximate solution~\cite{caputo1967linear}. 
In heat transfer and diffusion, time-space fractional diffusion equation models are studied in~\cite{atangana2016new,lin2021irk,mahor2021analytical}. 
Another application area of fractional calculus is structural engineering. 
One of the recent studies investigates the usage of fractional calculus for damping and hybrid base isolation system~\cite{li2020fractional}. 
Other applications are also present in the literature, for example for 
PID controllers~\cite{verma2021water}, 
acoustic waves in complex media~\cite{qu2020homotopy,jleli2020solution,hodaei2020transient}, 
wave equation~\cite{li2021mass,khan2019analytical}, 
nonlinear Hilfer fractional stochastic differential systems~\cite{sathiyaraj2019ulam},
epidemic models~\cite{baba2019existence,qureshi2019fractional,saratha2021analysis}, 
disease transmission model~\cite{yusuf2021mathematical}, 
prey-predator model~\cite{abbagari2021analytical}, 
quantum Stirling~\cite{aydiner2021space} and Szilard~\cite{aydiner2021quantum} heat engines, and financial systems with time delay~\cite{wang2020chaos}. 

Besides the applications of fractional derivatives to diverse branches of science, there exist many studies that promote various methods to develop such derivatives. 
The linear multi-term Caputo-type fractional equation with a spectral collocation method is studied in~\cite{cai2021fractional}. 
The fractional Laplacian is used as the space operator in a study of stochastic space fractional wave equation in~\cite{liu2021higher}, where the effects of additive noise term with an infinite or fractional Brownian motion is also considered. 
The criteria for the existence of solutions for a new fractional boundary value problem in the Liouville-Caputo setting by using Kuratowski’s measure of noncompactness, Sadowski’s fixed point theorem, Gronwall inequality, and the endpoint theory are studied in~\cite{mohammadi2021criteria}. 
The relations between the proportional Caputo derivative and the special function such as beta function~\cite{akgul2021analysis}, are also studied in the literature. This list is only a tiny fraction of the studies and applications of the fractional calculus, and can only serve as a brief introduction. The reader is referred to the aforementioned publications and references therein for a more comprehensive discussion.

The applications of fractional calculus to the Schr\"{o}dinger equation, which serves as a model in various fields including but is not limited to quantum physics, optics, hydrodynamics, and elastic body dynamics, are also quite many. 
The fractional Schr\"{o}dinger equation was first derived in~\cite{laskin2002fractional} and since then its many possible applications are proposed and studied. 
A Galerkin-Legendre spectral scheme that satisfies mass and energy conservation is studied in~\cite{fei2020linearized}. 
The contraction mapping principle is applied and the wave functions with their energy levels are studied in~\cite{shao2021cauchy}. The existence of the ground states for asymptotically linear fractional Schr\"{o}dinger equation by variational methods is proven in~\cite{chen2021ground}. 
Fractional Schr\"{o}dinger-Poisson system of equations is also studied and it is shown that ground state solution and a sign-changing solution exist for such a system~\cite{gu2021multiple}. 
In a semi-classical analysis of the fractional Schr\"{o}dinger equation, it is shown that the non-negative continuous potential can decay arbitrarily~\cite{an2021semi}. 
Various studies on fractional Schr\"{o}dinger equation is still the subject of many  ongoing studies~\cite{laskin2002fractional,jamshir2021time,alouini2022finite,liemert2016fractional}.

On the other hand, the Petviashvili method (PM) is one of the commonly used methods for obtaining the soliton solutions of nonlinear systems. 
PM was first introduced in~\cite{petviashvili1976equation}, and later is extended to spectral renormalization method~\cite{ablowitz2005spectral} to serve as a numerical scheme that can be applied to many nonlinear systems including nonlinearities not necessarily restricted to fixed homogeneities. 
To the best of our knowledge, the extension of PM to the fNLSE remains an open research problem. Additionally, the analysis of the interaction of the soliton solutions of the fNLSE with various forms of potentials also remains to be investigated. With these motivations, we propose a numerical PM for the construction of soliton solutions of the fNLSE. We construct one and two soliton solutions of the fNLSE using this numerical scheme and discuss their properties and the effects of fractional order derivative $\alpha$ on these solitons' characteristics and dynamics. 
q-deformed algebras~\cite{bonatsos2002deformed,altintas2011inhomogeneous,altintas2012inhomogeneous} and potentials~\cite{molaee2012s,lutfuouglu2018analytical} have been considered in various areas from relavistic wave equations~\cite{brzo2021klein}, quantum computing~\cite{altintas2014constructing,altintas2020q} and optomechanical systems~\cite{kundu2022transparency} to quantum metrology~\cite{hasegawa2006quantum}, quantum thermodynamics~\cite{guvenilir2022work} and so on. 
Therefore, in addition to zero potential and photorefractive potential, we consider q-deformed Rosen-Morse potentials and discuss their effects on soliton behavior and dynamics. 
We discuss our findings and their possible usage areas in applied sciences. 

\section{Mathematical Formulation of the Petviashvili's Method for the Solution of the Fractional Nonlinear Schr\"{o}dinger Equation}
Fractional order Schr\"{o}dinger equation was first derived in~\cite{laskin2002fractional} and became a common tool for modeling various processes in quantum mechanics, optics, hydrodynamics, just to name a few areas of its applications. The fractional order Schr\"{o}dinger can be written as 
\begin{equation}
	i\hbar \frac{\partial \psi(\mathbf{r},t)}{\partial t}=H_\alpha \psi(\mathbf{r},t),
	\label{eq01}
\end{equation}
where the dependent variable $\psi$ denotes the complex wavefunction, $i$ is the imaginary unity, $\hbar$ denotes the reduced Planck's constant, $\mathbf{r}$ is the position vector, $t$ is the time and the Hamiltonian is given by $H_\alpha=D_{\alpha}(-\hbar^2 \Delta)^{\frac{\alpha}{2}}+V(\mathbf{r},t)$ where $D_{\alpha}$ is a constant in dimensions of $energy^{1-\alpha}length^{\alpha}time^{-\alpha}$ which reduces to $D_{\alpha}=1/2m$ for $\alpha=2$ where $m$ denotes the mass of atomic particle, $\Delta=\partial^2 / \partial \mathbf{r}^2$ is the Laplacian, $V(\mathbf{r},t)$ is the potential function and $\alpha$ shows the order of the fractional derivative. Thus, in the full form the fractional Schr\"{o}dinger equation can be written as
\begin{equation}
	i\hbar \frac{\partial \psi(\mathbf{r},t)}{\partial t}=D_{\alpha}(-\hbar^2 \Delta)^{\frac{\alpha}{2}} \psi(\mathbf{r},t)+V(\mathbf{r},t)\psi(\mathbf{r},t).
	\label{eq02}
\end{equation}
Here, the 3D Riesz fractional derivative can be written as
\begin{equation}\label{eq03}
	(-\hbar^2 \Delta)^{\frac{\alpha}{2}} \psi(\mathbf{r},t)=\frac{1}{(2\pi \hbar)^3} \int d^3pe^{\frac{i}{\hbar}\mathbf{p}.\mathbf{r}}|\mathbf{p}|^{\alpha}\varphi(\mathbf{p},t),
\end{equation}
where $ \varphi(\mathbf{p},t)$ is the 3D Fourier transform of the wavefunction $\psi(\mathbf{r},t)$. In 1D and dimensionless form, the governing fractional Schr\"{o}dinger equation given in Eq.~(\ref{eq01}) reduces to~\cite{liemert2016fractional} 
\begin{equation}
	i \psi_{t}(x,t)=\frac{1}{2}(-\Delta)^{\frac{\alpha}{2}} \psi(x,t)+V(x,t)\psi(x,t).
	\label{eq05}
\end{equation}
We confine ourselves with time-independent potentials, $V(x)$ and we also take the effect of nonlinearity on the wavefunction, thus the 1D fractional nonlinear Schr\"{o}dinger equation (fNLSE) becomes
\begin{equation}
	i \psi_{t}(x,t)-\frac{1}{2}(-\Delta)^{\frac{\alpha}{2}} \psi(x,t)-V(x)\psi(x,t)+\sigma \left| \psi(x,t) \right|^2 \psi=0.
	\label{eq06}
\end{equation}
Here, $\sigma$ is a constant controlling the strength of the nonlinearity and is taken to be $\sigma=1$ throughout this study in conjunction with the existing literature~\cite{bayindir2020rogueHarmonic,kivshar2001nonlinear,bayindir2018freezing,bayindir2022self}.

There are a few techniques in the literature for the construction of soliton solutions of nonlinear equations such as shooting, self-consistency, and relaxation methods~\cite{petviashvili1976equation, ablowitz2005spectral,bayindir2019self}. 
One of the commonly used methods with this aim is the Petviashvili's method (PM)~\cite{petviashvili1976equation}. 
PM relies on transforming the governing dynamic equation into the wavenumber domain via FFT routines. After such a transformation, starting from the initial conditions, the iterations are continued until a nonlinearity-dependent convergence criterion is satisfied\cite{petviashvili1976equation,ablowitz2005spectral,bayindir2018compressive,bayindir2021self}. 
An energy conservation principle is used to check the convergence of the scheme~\cite{petviashvili1976equation,ablowitz2005spectral}. 
The method summarized here is first proposed by Petviashvili and applied to the Kadomtsev-Petviashvili (KP) equation for the construction of its soliton solutions~\cite{petviashvili1976equation}. 
Later, PM is extended to account for nonlinearities not necessarily restricted to fixed homogeneity and named as the spectral renormalization method (SRM)~\cite{ablowitz2005spectral,bayindir2021self}.  
To obtain the soliton solutions of different systems with missing spectral data compressive spectral renormalization method (CSRM) is proposed by one of us in~\cite{bayindir2018compressive}, which relies on the compressive sensing algorithm. 
The literature which utilizes PM and SRM is vast. 
These methods are commonly used to analyze the characteristics and dynamics of various phenomena such as lattice vortices, dark and gray solitons, the effect of various types of nonlinearities, wave blocking \cite{ablowitz2005spectral,bayindir2021selfwaveblocking,fibich2015nonlinear}.  
The reader can refer to papers~\cite{petviashvili1976equation,ablowitz2005spectral,fibich2015nonlinear} and the references therein for more detailed discussions of PM and SRM and their applications.

In this part, we extend the PM method for the solution of the fNLSE given in Eq.~(\ref{eq06}). One can rewrite Eq.~(\ref{eq06}) as
\begin{equation}
	i\psi_t -\frac{1}{2}(-\Delta)^{\frac{\alpha}{2}} \psi(x,t)-V(x)\psi+ N(\left| \psi \right|^2) \psi =0,
	\label{eq07}
\end{equation}
where the nonlinear term reads as $N(\left| \psi \right|^2)=\sigma \left| \psi \right|^2$. Using the ansatz, $\psi(x,t)=\eta(x,\mu) \textnormal{exp}(i\mu t)$, where $\mu$ is the soliton eigenvalue, the fNLSE with becomes
\begin{equation}
	-\mu \eta  -\frac{1}{2}(-\Delta)^{\frac{\alpha}{2}}\eta -V(x)\eta+ N(\left| \eta \right|^2) \eta =0.
	\label{eq08}
\end{equation}
The Fourier transform of $\eta$ parameter can be found using
\begin{equation}
	\widehat{\eta} (k)=F[\eta(x)] = \int_{-\infty}^{+\infty} \eta(x) \exp[i(kx)]dx,
	\label{eq09}
\end{equation}
where $F$ denotes the Fourier transform. Thus, the Fourier transform of the fractional order derivative in Eq.~(\ref{eq07}) can be computed using
\begin{equation}\label{eq10}
	F[(-\Delta)^{\frac{\alpha}{2}} \eta(x)] = \left| k \right|^\alpha \widehat{\eta} (k),
\end{equation}
where the parameter $k$ is the wavenumber array which has $N$ multiples of the fundamental wavenumber, $k_o=2 \pi/L$ as its entries. Fourier inverting the expression in Eq.(\ref{eq10}) via IFFTs, the fractional-order derivative can be computed~\cite{liemert2016fractional,bayindir2015compressive, trefethen2000spectral,bayindirMS,canuto2007spectral,karjadi2011effects}. 
The number of spectral components should be selected as a power of 2 for the efficient computations of FFT and IFFT routines. 

The iteration equation in the wavenumber domain can be obtained by taking the 1D Fourier transform of Eq.~(\ref{eq07}), however, it is well-known that such an iteration equation would have a singularity~\cite{petviashvili1976equation,ablowitz2005spectral}. 
This is also true for the fNLSE investigated in this paper. 
To avoid such a singularity we add and subtract a $p \eta$ term to the Eq.~(\ref{eq07}). 
Here, $p$ denotes a positive real number, which is selected to be $p=10$ throughout this study~\cite{petviashvili1976equation,ablowitz2005spectral}. 
After such an operation, the 1D Fourier transform of Eq.~(\ref{eq07}) yields 
\begin{equation}
	\widehat{\eta} (k)=\frac{(p+| \mu|)\widehat{\eta}}{p+0.5 \left| k \right|^\alpha} -\frac{F[V \eta]-F \left[ N(\left| \eta \right|^2) \eta \right]}{p+0.5 \left| k \right|^\alpha},
	\label{eq11}
\end{equation}
which becomes the PM iteration scheme for the fNLSE. Following other studies on PM and SRM, we introduce the parameter $\xi(x)$ defined by $\eta(x)=\gamma \xi(x)$ to obtain a convergence condition. The Fourier transform of this new parameter can be computed by $\widehat{\eta}(k)=\gamma \widehat{\xi}(k)$. With such a substitution, the iteration formula given for the fNLSE in Eq.~(\ref{eq11}) becomes
\begin{equation}
	\widehat{\xi}_{j+1} (k) =\frac{(p+| \mu|)}{p+0.5\left| k \right|^\alpha}\widehat{\xi_j}-\frac{F[V \xi_j]}{p+0.5\left| k \right|^\alpha} +  \frac{F\left[\sigma |\gamma_j|^2|\xi_j|^2 \xi_j \right] }{p+0.5\left| k \right|^\alpha} =R_{\gamma_j}[\widehat{\xi}_j (k)].
	\label{eq12}
\end{equation}
By multiplying the both sides of Eq.~(\ref{eq12}) with $\widehat{\xi}^*(k)$ term and integrating for the calculation of the total energy, the convergence criteria for the PM proposed for fNLSE becomes
\begin{equation}
	\int_{-\infty}^{+\infty} \left|\widehat{\xi} (k)\right|^2 dk= \int_{-\infty}^{+\infty} \widehat{\xi}^* (k) R_{\gamma}[\widehat{\xi} (k)]dk.
	\label{eq13}
\end{equation}
The energy terms in the Eq.~(\ref{eq13}) can be calculated via ordinary summation operations in the wavenumber domain, which generally yields to different values. A parameter $\gamma$ can be introduced into this equation, and the ratio of the energy terms from the two sides of the equation can be obtained. Thus, the convergence of the scheme can be tested using the $\gamma_j$ parameter calculated at every time step, enabling the construction of converged soliton solutions of the fNLSE. Starting from the initial conditions described in terms of Gaussians, Eq.~(\ref{eq12}) and Eq.~(\ref{eq13}) are applied iteratively, until the parameter ${\gamma}$ convergences to a relative error bound which is selected as $10^{-4}$ in this study.

PM is a powerful method for the efficient computation of the soliton solutions of the various dynamic equations. 
However, the stability of the solitons obtained via PM is not guaranteed, thus it is necessary to analyze their stability dynamics. 
It is well-known that a necessary, but not a sufficient condition, for the soliton stability is the Vakhitov-Kolokolov slope condition~\cite{vakhitov1973stationary}. 
Vakhitov-Kolokolov slope condition dictates that for a soliton to be stable, the $dP/d \mu >0$ condition must be satisfied for the positive values of the soliton eigenvalues, $\mu>0$. 
In this formula, the soliton power parameter can be calculated using $P=\int |\psi|^2dx$. 
The second condition for the soliton stability is the spectral condition, which is introduced and discussed analytically in~\cite{weinstein1985modulational,sivan2008qualitative}. 
Another approach commonly used to test the spectral condition is the numerical approach. 
The complexity of the different equations involving complicated potential functions and defect modes make the numerical approach a popular choice. 
When the spectral condition is tested numerically, starting from the soliton solutions constructed by PM the time-stepping is performed for the next times using a time integration technique. 
With this aim, we implement a Fourier spectral scheme for the solution of the fNLSE. 
For the time-stepping, we use a $4^{th}$ order Runge-Kutta time integrator. 
For this purpose, the fNLSE presented in Eq.(\ref{eq06}) can be rewritten as
\begin{equation}
	\psi_t =i \left( -\frac{1}{2}(-\Delta)^{\frac{\alpha}{2}} \psi(x,t) -V(x) \psi + \sigma  \left|\psi \right|^2 \psi \right) =g(\psi, t,x).
	\label{eq14}
\end{equation}
Here, the function $g(\psi, t,x)$ denotes the right-hand-side of the expression given in Eq.(\ref{eq14}). The fractional order Laplacian in Eq.(\ref{eq14}) can be computed using the expression given in Eq.~(\ref{eq10}) and IFFT routines. All products in Eq.(\ref{eq14}) are calculated by ordinary multiplication in the physical domain. Time stepping of this equation is performed using a $4^{th}$ order Runge-Kutta algorithm. The four slopes of the Runge-Kutta at each time step are computed by
\begin{equation}
	\begin{split}
		& s_1=g(\psi^n, t^n, x) \\
		& s_2=g(\psi^n+0.5 s_1dt, t^n+0.5dt, x) \\
		& s_3=g(\psi^n+0.5 s_2dt, t^n+0.5dt, x) \\
		& s_4=g(\psi^n+s_3dt, t^n+dt, x), \\
	\end{split}
	\label{eq15}
\end{equation}

\noindent where $n, dt$ are the time index and the time step, respectively. The dimensionless time step is selected to be $dt=5 \times 10^{-3}$ throughout this paper, which satisfies the stability and convergence conditions. Starting from the soliton solutions of the fNLSE obtained by the PM, the complex wavefunction and time parameter at later time steps are found by
\begin{equation}
	\begin{split}
		& \psi^{n+1}=\psi^{n}+dt(s_1+2s_2+2s_3+s_4)/6, \\
		& t^{n+1}=t^n+dt.\\
	\end{split}
	\label{eq16}
\end{equation}
At every time step, the soliton power, $P$ is computed and time versus soliton power, $t-P$, graphs are plotted for the investigation of the stability characteristics of the solitons of the fNLSE. Additionally, the effect of potential terms on the characteristics and dynamics of these solitons are also investigated. The potentials considered for this purpose are photorefractive and q-deformed Rosen-Morse potentials, as described in the coming section of this paper.

\section{Results and Discussion}
\subsection{Results for Zero Potential, $V=0$}

\begin{figure}[t!]
	\begin{center}
			\hspace*{-0.1\columnwidth}
		\includegraphics[width=1.2\columnwidth]{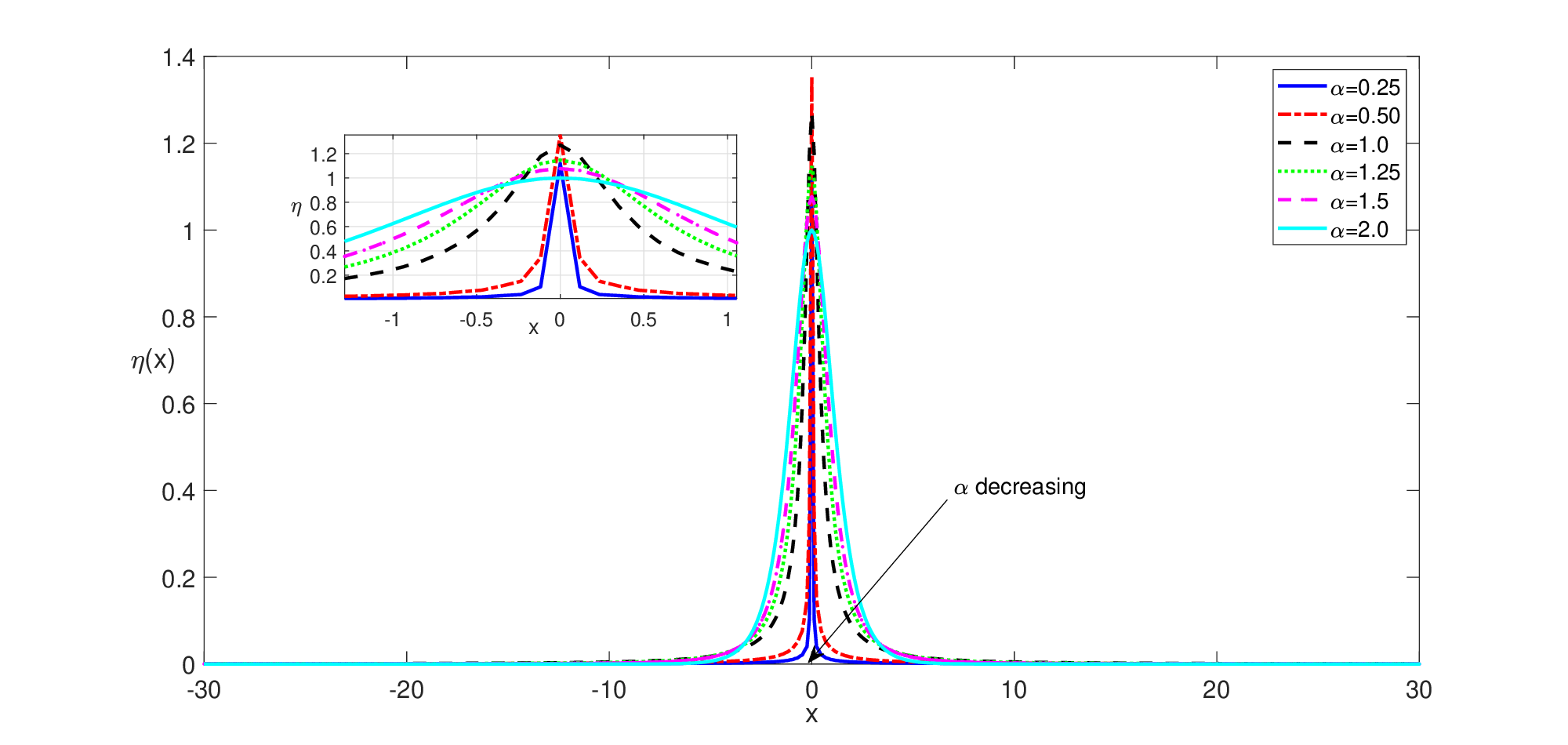}
	\end{center}
	\caption{\small Self-localized one soliton solutions of the fNLSE for various values of $\alpha$ for a zero potential of $V=0$.}
	\label{fig1}
\end{figure}

In this section, we present the characteristics and dynamics of the soliton solution of the fNLSE under the effect of zero optical potential, $V=0$. Starting from a Gaussian in the form of $\exp{(-x^2)}$ initially, the one soliton solution of the fNLSE can be obtained numerically using the PM method suggested above for various values of the fractional order, $\alpha$. The computational domain is set to be $x=[-L/2,L/2]$ where the value of $L=120$ is used for the length of the computational domain. The number of spectral components for this and all other simulations are selected as $N=2048$ for utilizing the full computational efficiency of the FFT and IFFT algorithms. The one soliton solution of the fNLSE obtained by the proposed  PM is shown in Fig.~\ref{fig1} for various values of fractional order, $\alpha$. 

In this figure, the amplitude of the one soliton solution of the fNLSE obtained for the fractional-order $\alpha=2$ is used as the normalization amplitude. Fig.~\ref{fig1} confirms that a decrease in the fractional-order, $\alpha$, leads to significant changes in the soliton profile. The solitons obtained for smaller $\alpha$ values are more peaked and slender, leading to a higher nonlinearity parameter. This increased slenderness behavior and significantly reduced soliton width can lead to many important applications in applied sciences. In nonlinear and fiber optics, and in hydrodynamics, the success of pulse/wave positioning, pinpointing, and matched filtering can be enhanced by adjusting the order of the fractional derivative, $\alpha$. Additionally, the increased slenderness of the solitons enhances their sparsity in the spatial and temporal domains. Higher sparsity of the soliton signals makes the usage of smart algorithms such as compressive sensing possible, bringing up many challenges. Efficient analysis, measurement, interpolation, and extrapolation of solitons of the fNLSE are a few of the problems involving those challenges.

\begin{figure}[t!]
	\begin{center}
		\hspace*{-0.1\columnwidth}
		\includegraphics[width=1.2\columnwidth]{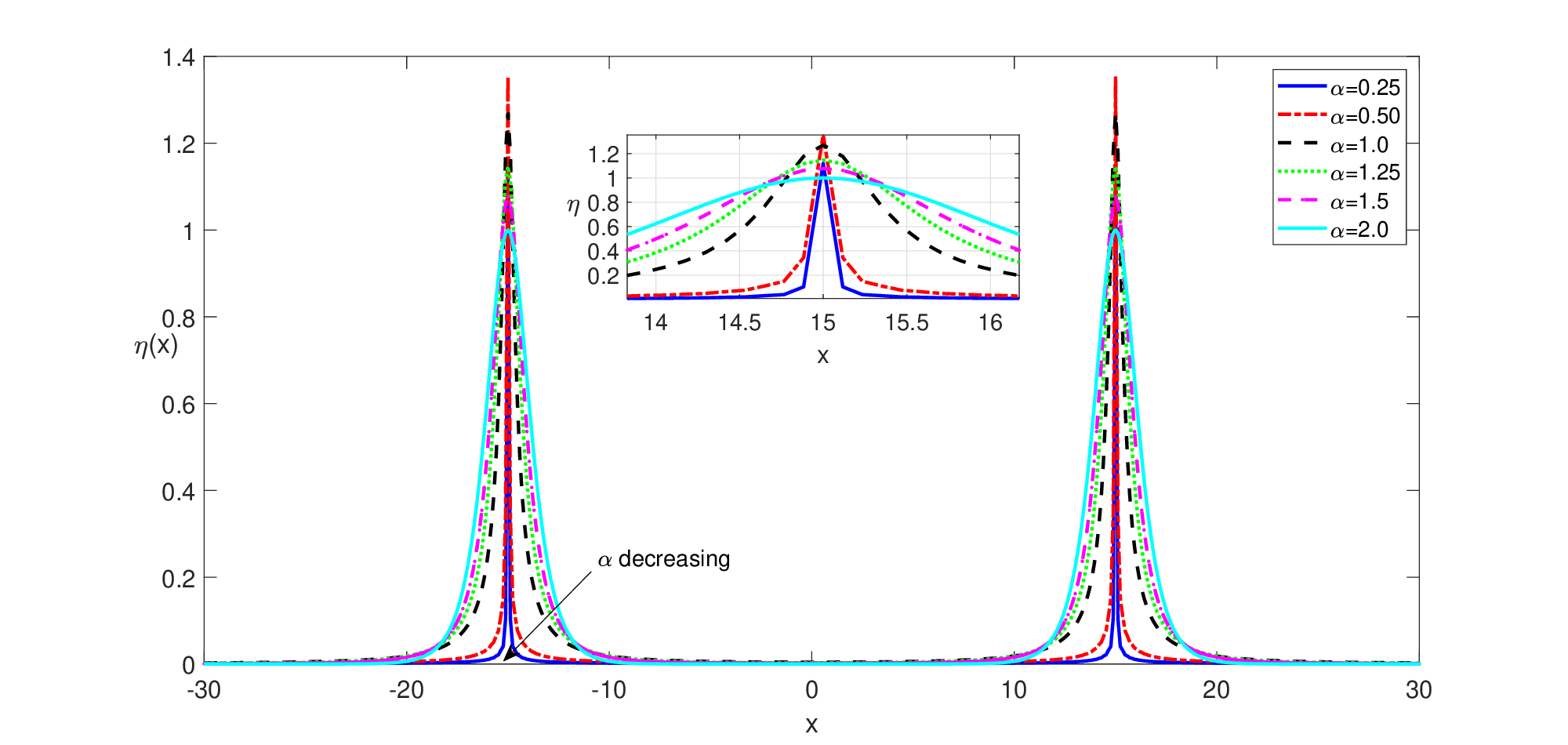}
	\end{center}
	\caption{\small Self-localized two soliton solutions of the fNLSE for various values of $\alpha$ for a zero potential of $V=0$.}
	\label{fig2}
\end{figure}

When the aforementioned computational approach is initiated using two Gaussians in the form of $\exp{(-(x-15)^2)}+\exp{(-(x+15)^2)}$, the PM converges to self-localized two soliton solutions of the fNLSE. Two solitons of the fNLSE obtained this way are displayed in Fig.~\ref{fig2} for various values of the fractional order, $\alpha$. Similar to the one soliton solutions, the two solitons presented in Fig.~\ref{fig2} become more slender as the fractional order parameter,  $\alpha$, decreases.

\begin{figure}[t!]
	\begin{center}
		\hspace*{-0.1\columnwidth}
		\includegraphics[width=1.2\columnwidth]{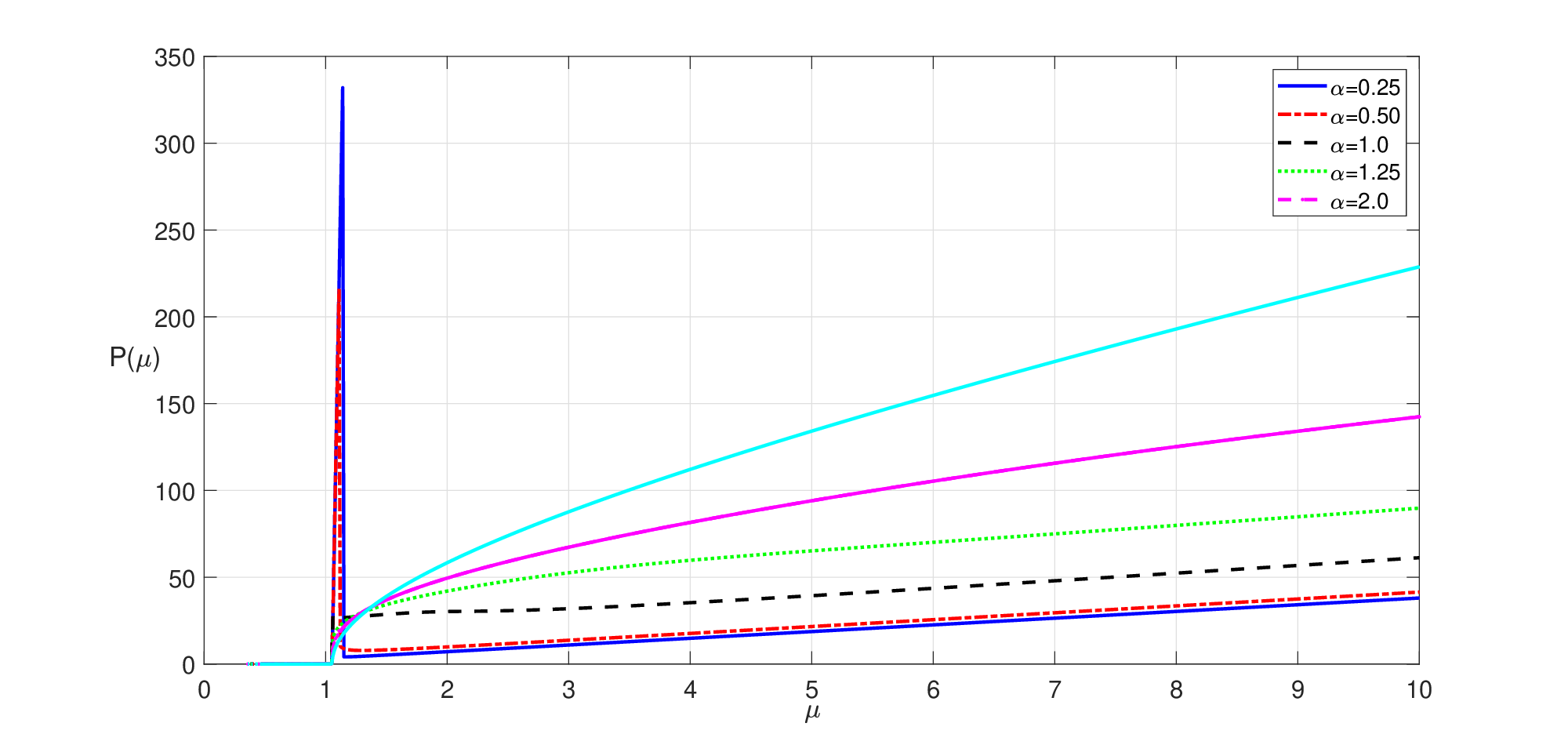}
	\end{center}
	\caption{\small Self-localized one soliton power vs soliton eigenvalue, $\mu$, for various values of $\alpha$ for a zero potential of $V=0$.}
	\label{fig3}
\end{figure}

\begin{figure}[t!]
	\begin{center}
		\hspace*{-0.1\columnwidth}
		\includegraphics[width=1.2\columnwidth]{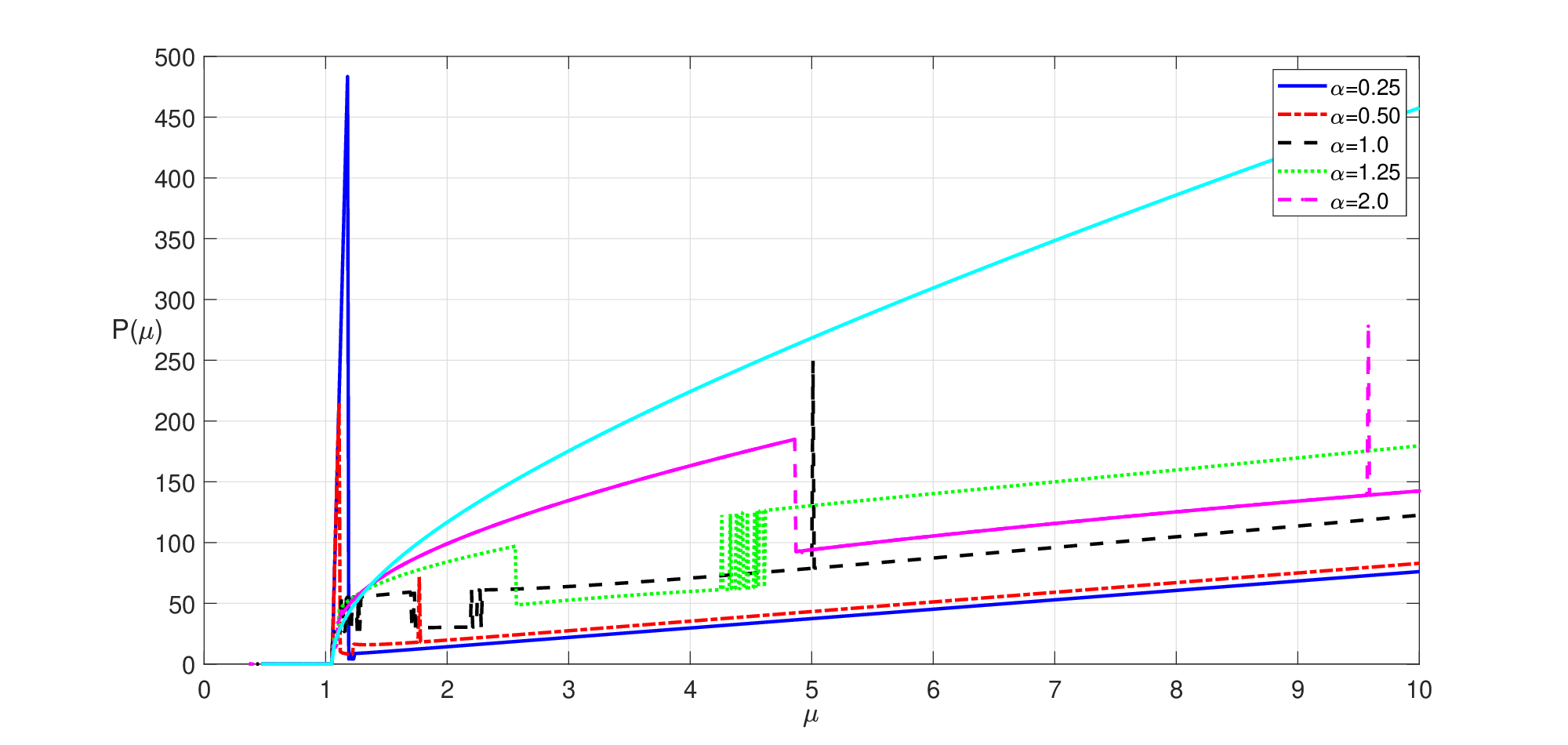}
	\end{center}
	\caption{\small Self-localized two soliton power vs soliton eigenvalue, $\mu$, for various values of $\alpha$ for a zero potential of $V=0$.}
	\label{fig4}
\end{figure}

Although the solitons of the fNLSE with 1,2, or multi-peaks can be constructed using the PM method summarized above, their stability characteristics should also be investigated. Due to this reason, we refer to some studies in the literature which investigate the stability characteristics of solitons (see e.g.~\cite{vakhitov1973stationary,weinstein1985modulational,sivan2008qualitative} and the references therein). Vakhitov-Kolokolov condition, also known as the slope condition~\cite{vakhitov1973stationary}, is one of the necessary conditions to be satisfied for the soliton stability. 
This condition dictates that for the soliton to be stable, the $dP/d \mu >0$ condition should hold true for the positive values of the soliton eigenvalue, $\mu>0$~\cite{vakhitov1973stationary}. 
Here, $P=\int |\psi|^2dx$ denotes the soliton power.
To examine if the Vakhitov-Kolokolov condition is satisfied, we depict the soliton eigenvalue vs power graphs in Fig.~\ref{fig3} and Fig.~\ref{fig4} for the one soliton and two soliton solutions of the fNLSE, respectively. For the range of soliton eigenvalues considered, which is $\mu \in [0,10]$, we observe that the Vakhitov-Kolokolov condition is satisfied with the exceptional regions and spikes in Fig.~\ref{fig3} and \ref{fig4}. The behavior and trend in the graphs are quite similar for the different values of the derivative fractional order, $\alpha$. Especially the $\mu-P$ graph for the one soliton case exhibits a smoother behavior. Since for the soliton eigenvalue of $\mu=2$ the Vakhitov-Kolokolov is $dP/d \mu >0$ satisfied, the value of $\mu=2$ is used for computations presented in this paper.

\begin{figure}[t!]
	\begin{center}
		\hspace*{-0.1\columnwidth}
		\includegraphics[width=1.2\columnwidth]{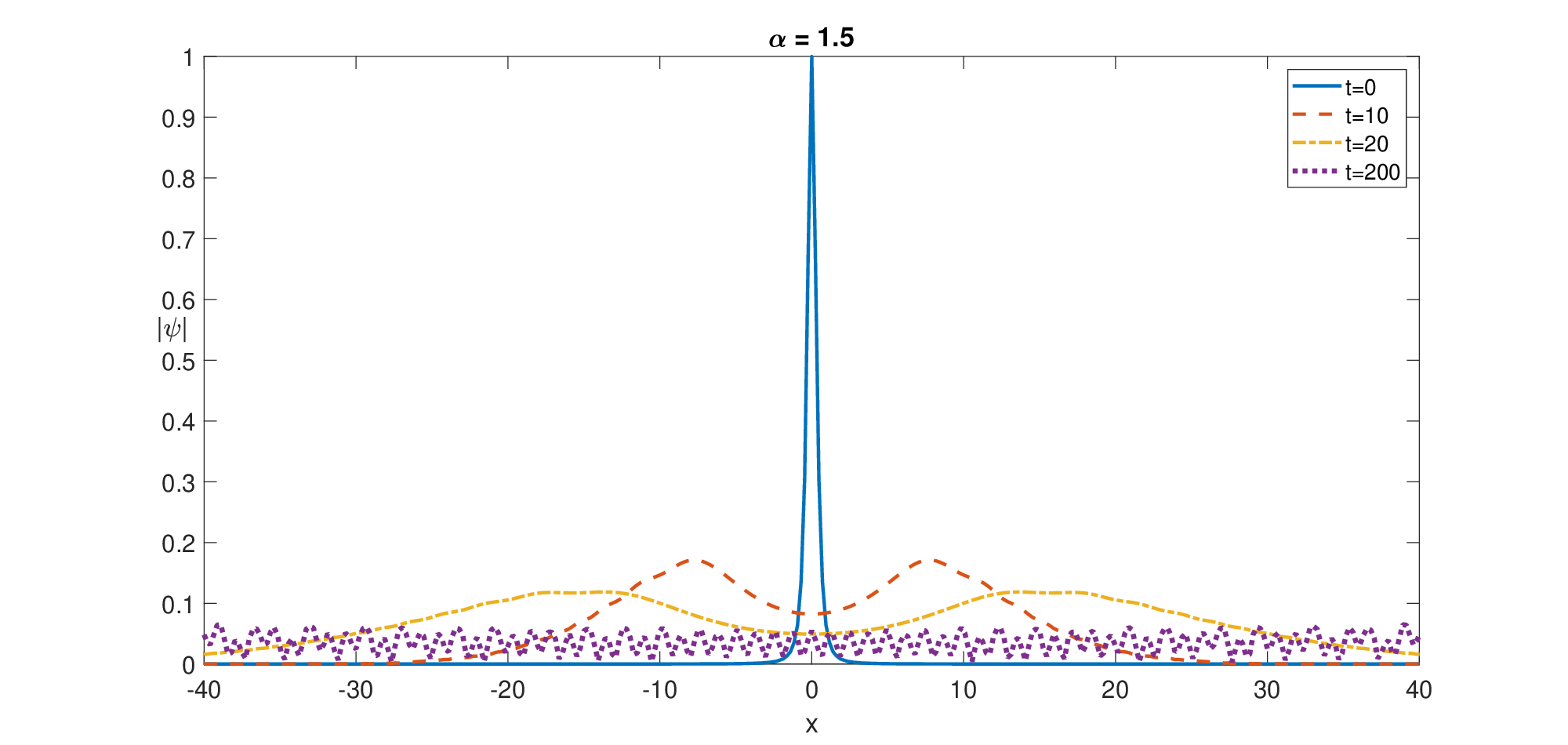}
	\end{center}
	\caption{\small Self-localized one soliton profiles at different times for $\alpha=1.5$ for a zero potential of $V=0$.}
	\label{fig5}
\end{figure}

\begin{figure}[t!]
	\begin{center}
		\hspace*{-0.1\columnwidth}
		\includegraphics[width=1.2\columnwidth]{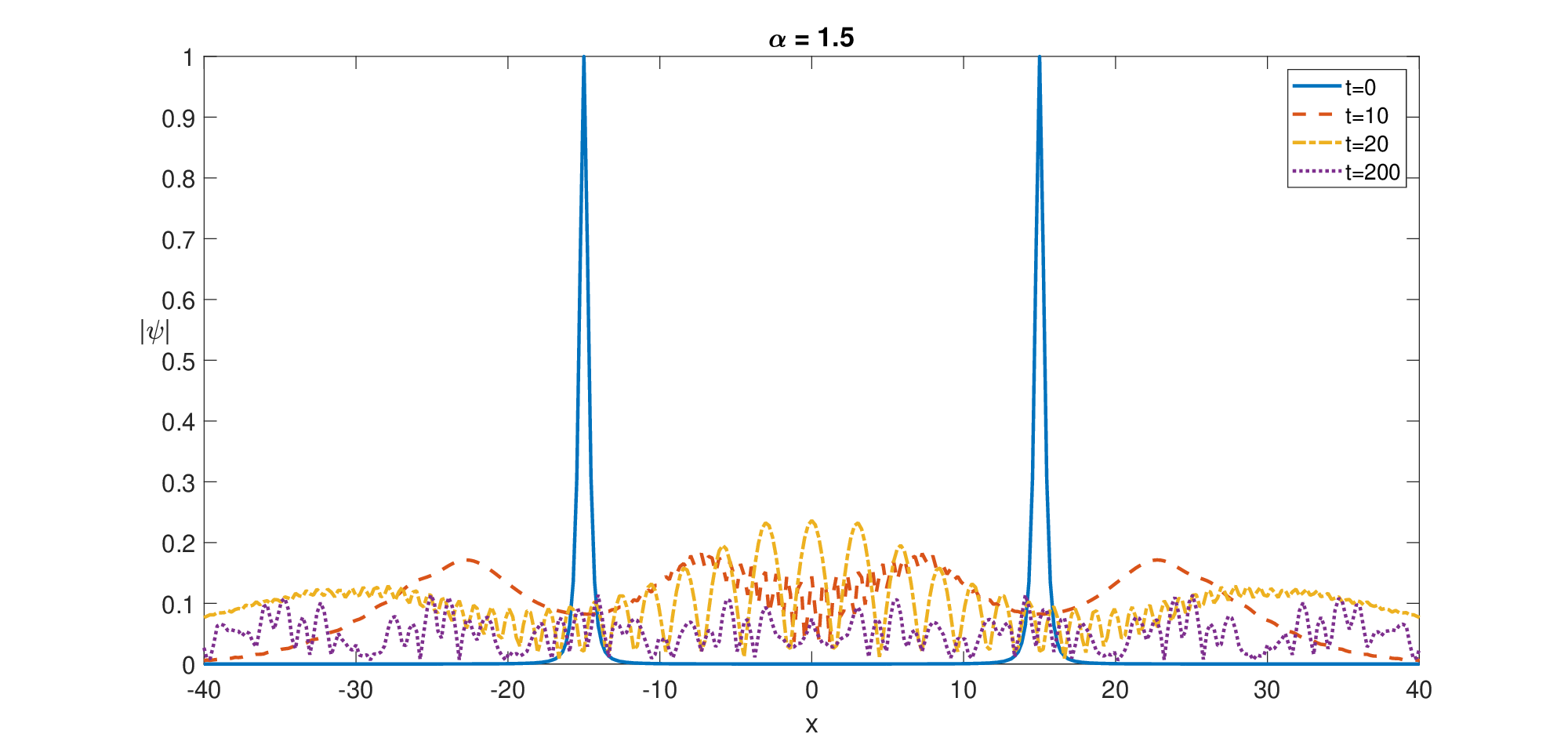}
	\end{center}
	\caption{\small Self-localized two soliton profiles at different times for $\alpha=1.5$ for a zero potential of $V=0$.}
	\label{fig6}
\end{figure}

In the soliton stability theory, it is known that the Vakhitov-Kolokolov slope condition is necessary for the soliton stability, however, it is not a sufficient condition. Thus, the spectral condition should be investigated either analytically or numerically~\cite{weinstein1985modulational,sivan2008qualitative}. 
For this purpose, we use the spectral method with the $4^{th}$ order Runge-Kutta time-stepping algorithm and analyze the temporal soliton characteristics and profiles numerically starting from the soliton shapes obtained by the PM method proposed for fNLSE and presented in Figs.~\ref{fig1} and \ref{fig2}. We depict the one soliton profile at different times in Fig.~\ref{fig5} and two soliton profiles at different times in Fig.~\ref{fig6}. For both of these figures, the derivative fractional order is selected as $\alpha=1.5$.

Checking Figs.~\ref{fig5} and Figs.~\ref{fig6}, we can observe both of the one and two soliton profiles are exhibiting a splitting and spreading behavior. Such a splitting and spreading behavior is observed for all derivative fractional orders in the interval of $0<\alpha<2$. In the long-time limit, both the one and two soliton profiles evolve into sinusoidal-looking waveforms. This splitting and spreading dynamics is also commensurate with the findings presented in~\cite{liemert2016fractional}. 
For the fractional order of $\alpha=2$, well-defined initial soliton shapes are preserved during the temporal evolution as expected, since the governing fNLSE reduces to the NLSE for this fractional derivative value.

\begin{figure}[t!]
	\begin{center}
		\hspace*{-0.1\columnwidth}
		\includegraphics[width=1.2\columnwidth]{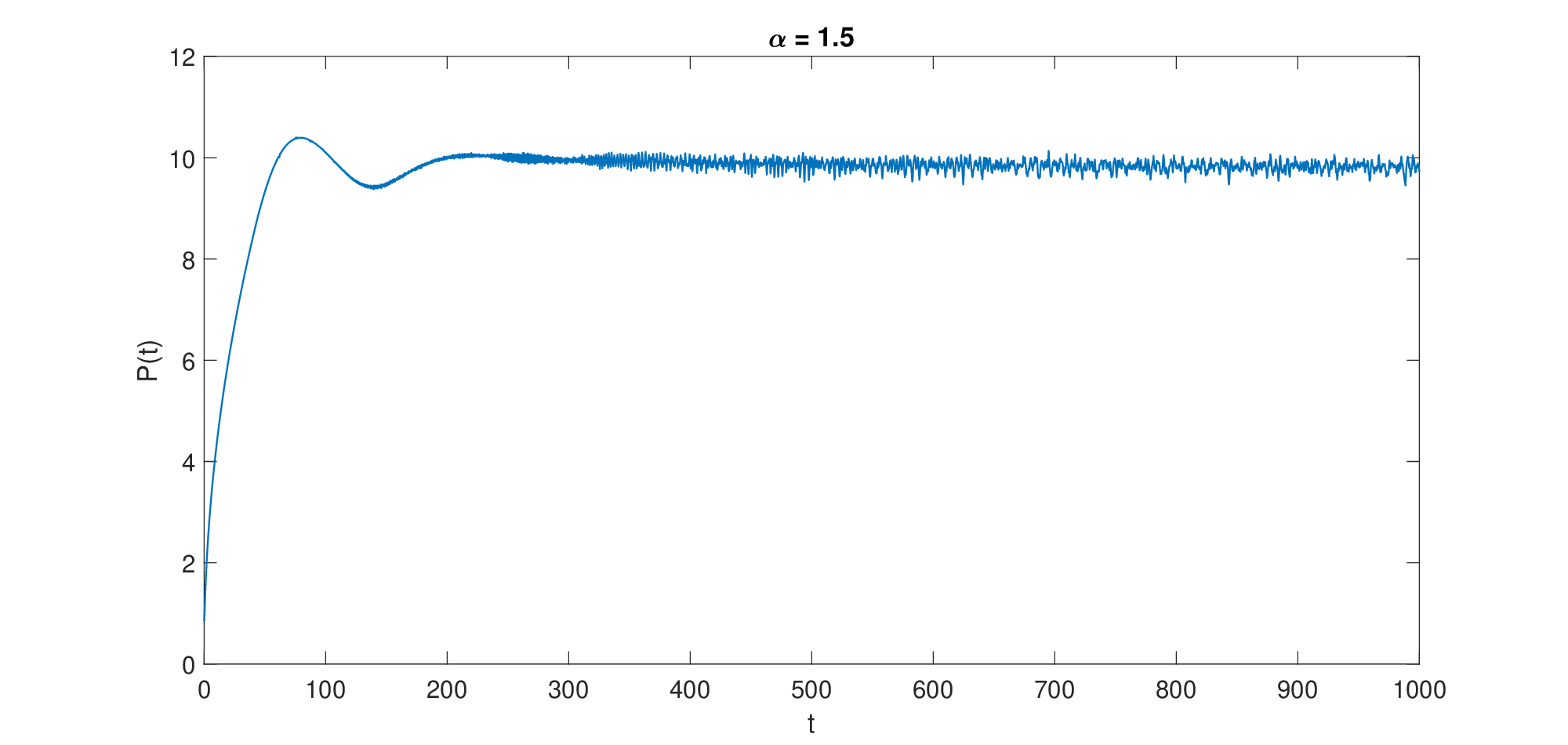}
	\end{center}
	\caption{\small Self-localized one soliton power as a function of time for $\alpha$ for a zero potential of $V=0$.}
	\label{fig7}
\end{figure}

In order to examine the stabilities or instability characteristics of the solitons if any, we plot the time vs soliton power graph in Fig.~\ref{fig7}. Since both of the one and two soliton $t-P$ graphs give similar results, only the graph for the one soliton dynamics is presented in Fig.~\ref{fig7}. Checking this figure, it is possible to conclude that soliton power gets stabilized after some adjustment time and in the long-time limit exhibit a steady-in-the-mean behavior. The low amplitude oscillations observed in the long time limit in the $t-P$ graphs presented in Fig.~\ref{fig7} are due to the last forms of the profiles exhibiting a sinusoidal-like behavior. We observe similar results for different values of $0<\alpha<2$, however for the sake of brevity and clarity of the presentation only the $t-P$ graph for $\alpha=1.5$ is presented.

\begin{figure}[t!]
	\begin{center}
		\hspace*{-0.1\columnwidth}
		\includegraphics[width=1.2\columnwidth]{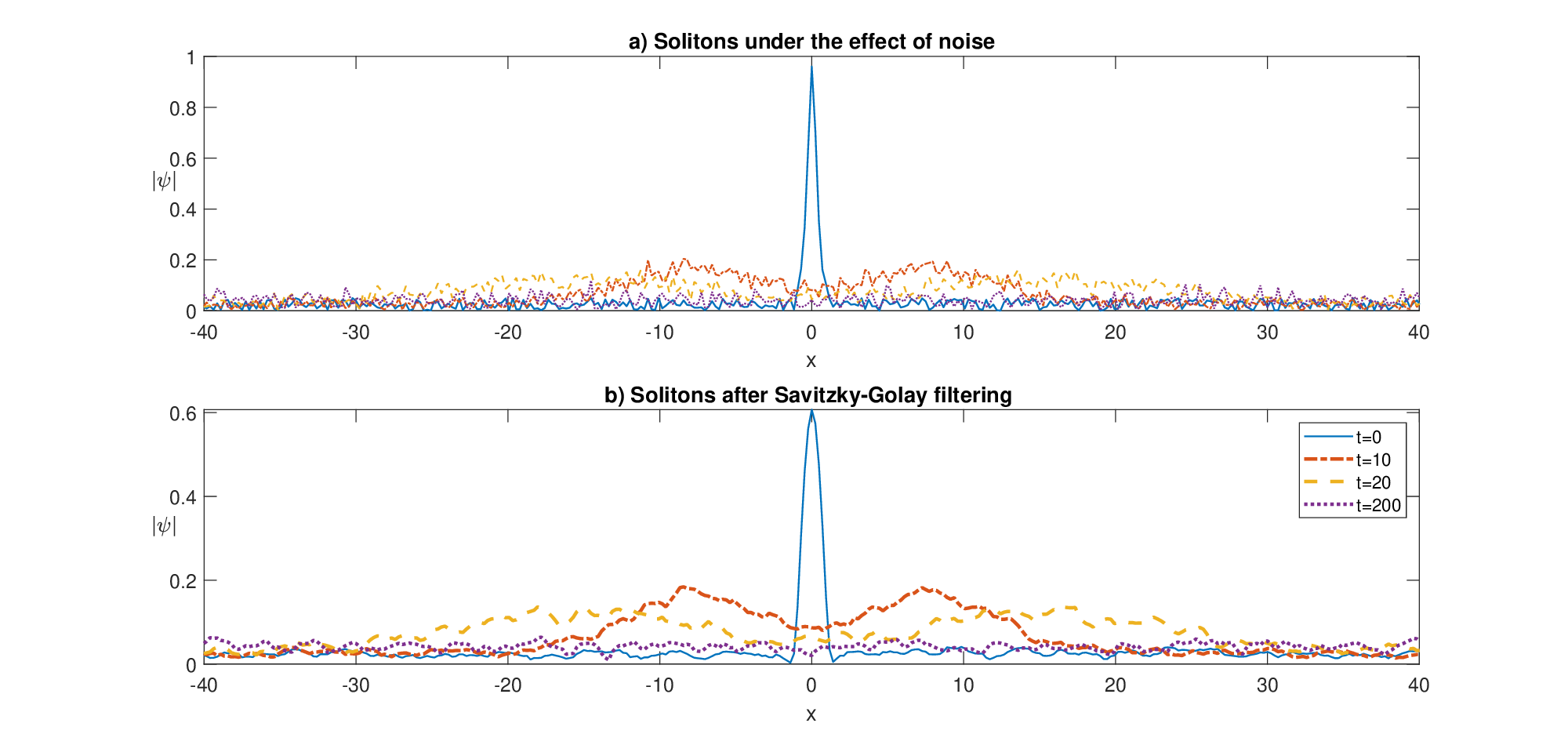}
	\end{center}
	\caption{\small Self-localized one soliton profiles at different times for $\alpha=1.5$ for a zero potential of $V=0$. a) Under the effect of noise b) After the noise is filtered out by Savitzky-Golay filter.}
	\label{fig8}
\end{figure}

Next, we test the robustness of the soliton solutions against the noisy perturbations. With this motivation, we impose a white noise in the form $\epsilon=\kappa a(x)$ to the wavefunction at every time step of temporal evolution, where the noise terms are $\kappa=0.05$ and $a(x)$ includes uniformly distributed random numbers in the interval of $[-1,1]$. As Fig.~\ref{fig8} confirms, the soliton solutions of the fNLSE are robust against such perturbations and they remain bounded. No rogue wave patterns triggered by modulational instability is observed, including the long-time limiting case at which the profile resembles a sinusoidal waveform. In Fig.~\ref{fig8}, in the first subplot, the self-localized one soliton shape at different times under the effect of noisy perturbations is depicted. In the second subplot of Fig.~\ref{fig8}, the denoised soliton profiles at various times are presented. For the denoising process, we used a $3^{rd}$ order Savitzky-Golay filter. As the figure indicates, the denoising filter can be used to filter the noise to get the fundamental characteristics and profiles of the solutions, however, such a denoising process can cause some reduction in the peak amplitudes of the solitons, which become even more pronounced for the lower values of the fractional-order $\alpha$. Similar behavior is also observed for the two soliton case.

\subsection{Results for Photorefractive Potential, $V=I_o\cos^2(k_x x)$}
Various forms of the potential functions are commonly used to model the effects of different phenomena on nonlinear processes. In quantum mechanics parabolic traps~\cite{bayindir2020rogueHarmonic,kivshar2001nonlinear}, linear potentials~\cite{liemert2016fractional} and $q$-deformed potentials~\cite{bayindir2021self,molaee2012s,lutfuouglu2018analytical,hatami2019analytical,mcfarlane89,biedenharn89,biedenharn1995quantum} are well-studied forms of the potentials. 
In nonlinear and fiber optics, the photorefractive and saturable photorefractive potentials with/without defects are commonly investigated~\cite{ablowitz2005spectral,bayindir2019self,bayindir2021self,fibich2015nonlinear}. 
In hydrodynamics, the blocking effect of opposing currents is modeled using constant, linear, sinusoidal, and other types of potentials~\cite{bayindir2021selfwaveblocking,kharif2003physical}. 
In Bose-Einstein condensation, repulsive and attractive potentials are used to model ultracold atomic systems dynamics with optical traps~\cite{khaykovich2002formation,strecker2002formation,nguyen2017formation}. 
This brief list is intended to provide the reader a starting point, and by no means complete.

\begin{figure}[t!]
	\begin{center}
		\hspace*{-0.1\columnwidth}
		\includegraphics[width=1.2\columnwidth]{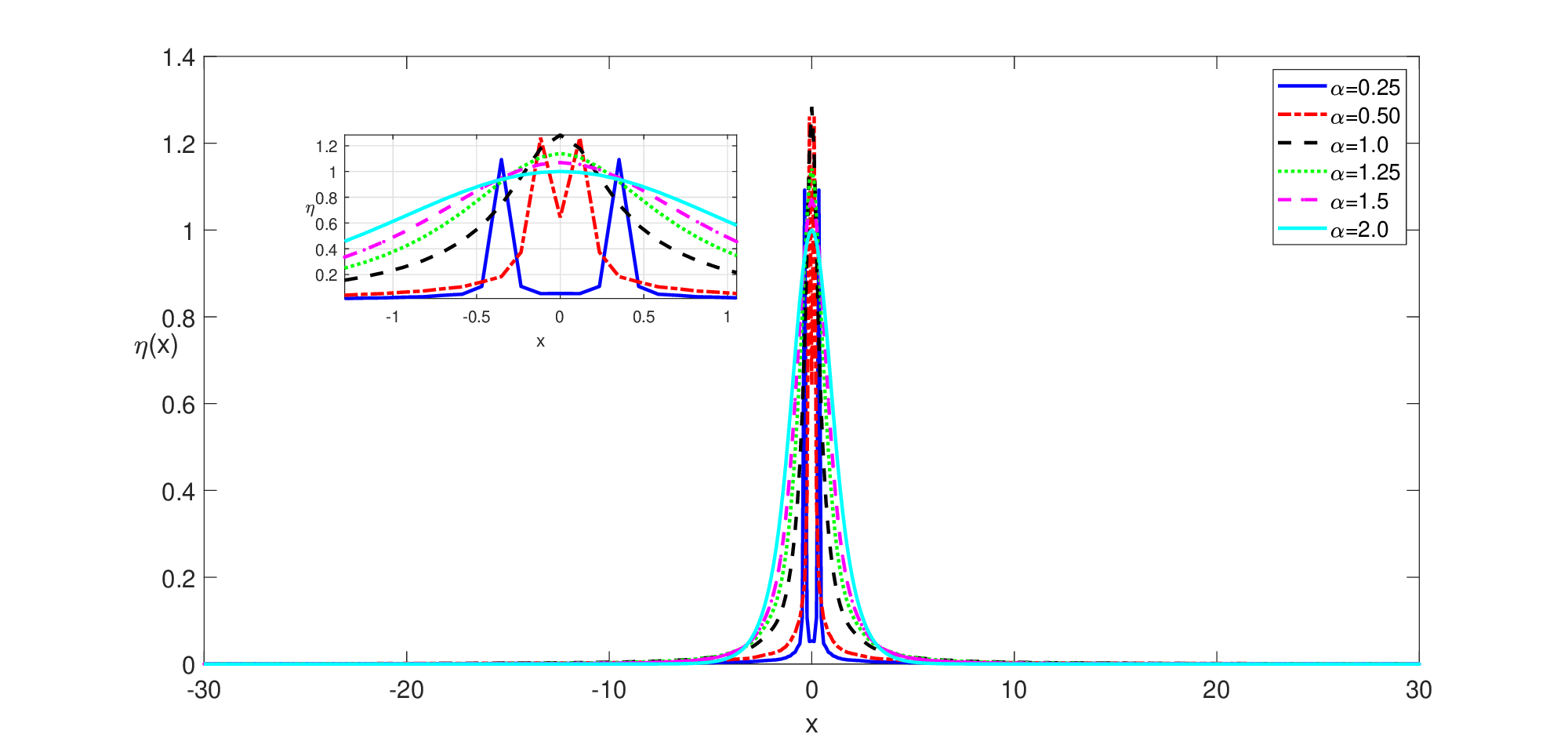}
	\end{center}
	\caption{\small Self-localized one soliton solutions of the NLSE for various values of $\alpha$ for a photorefractive potential of $V=I_o\cos^2(k_x x)=2 \cos^2(2x)$.}
	\label{fig9}
\end{figure}

In this section, we investigate the effects of the photorefractive potentials on the soliton solutions of the fNLSE. With this aim, we use a potential function in the form of $V=I_o\cos^2(k_x x)=2 \cos^2(2x)$, and follow the similar steps described above. Starting from a Gaussian initial condition, the one soliton solution of the fNLSE with the photorefractive potential is constructed and depicted in Fig.~\ref{fig9} for various values of the fractional order, $\alpha$.
Compared to the solitons depicted in Fig.~\ref{fig1}, the ones depicted in Fig.~\ref{fig9} have some differences. The striking effect of the photorefractive potential is that, although one soliton shape is preserved at the base of the soliton, some of them begin exhibiting two peaks. This finding is also confirmed by the results presented in~\cite{ablowitz2005spectral,bayindir2019self,bayindir2021self,fibich2015nonlinear}. 
These two peaked one soliton shapes become more prominent for the lower values of $\alpha$ such as $\alpha=0.25,0.50$ which lead to more slender profiles.

\begin{figure}[t!]
	\begin{center}
		\hspace*{-0.1\columnwidth}
		\includegraphics[width=1.2\columnwidth]{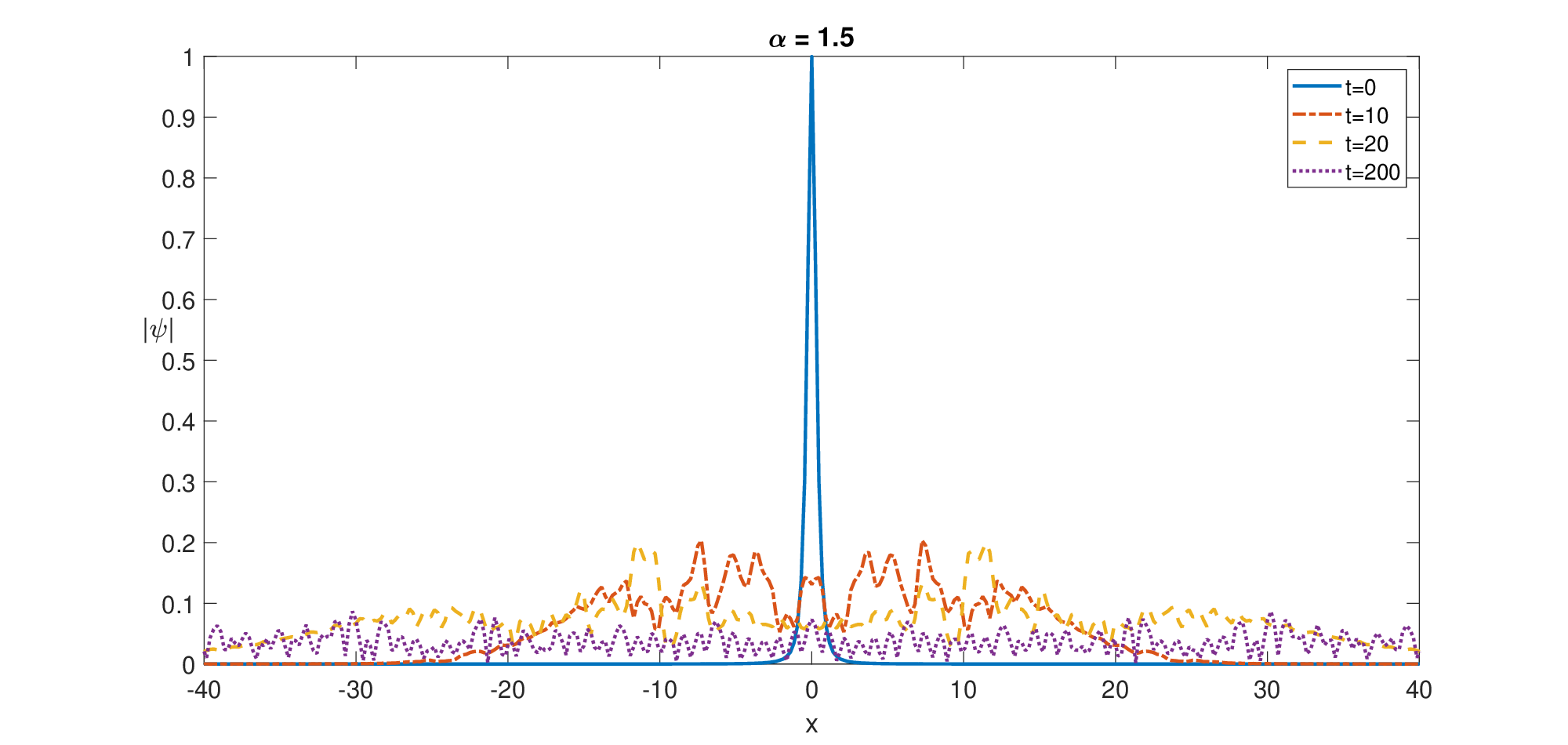}
	\end{center}
	\caption{\small Self-localized one soliton profiles at different times for $\alpha=1.5$ for a photorefractive potential of $V=I_o\cos^2(k_x x)=2\cos^2(2x)$.}
	\label{fig10}
\end{figure}
Next, we analyze the effect of the photorefractive potential $V=I_o\cos^2(k_x x)=2\cos^2(2x)$ on the temporal dynamics of the soliton shapes of the fNLSE. Running the time-stepping algorithm for the fNLSE for the derivative fractional order of $\alpha=1.5$ with this photorefractive potential, we obtain the soliton shapes which are depicted in Fig.~\ref{fig10}. As before, the splitting and spreading behavior for the solitons are experienced. However, a comparison of 
Figs.~\ref{fig5} and \ref{fig10} indicate that, the photorefractive potential causes the generation of higher wavenumber components appearing as shorter wavelength undulations on the main soliton profiles, which is especially clear for shorter times of evolution. Additionally, it is possible to conclude that photorefractive potential decreases the spreading of the one soliton. This is clear in Fig.~\ref{fig10}, since the peaks are mainly confined to a region closer to the origin $x=0$.

\begin{figure}[t!]
	\begin{center}
		\hspace*{-0.1\columnwidth}
		\includegraphics[width=1.2\columnwidth]{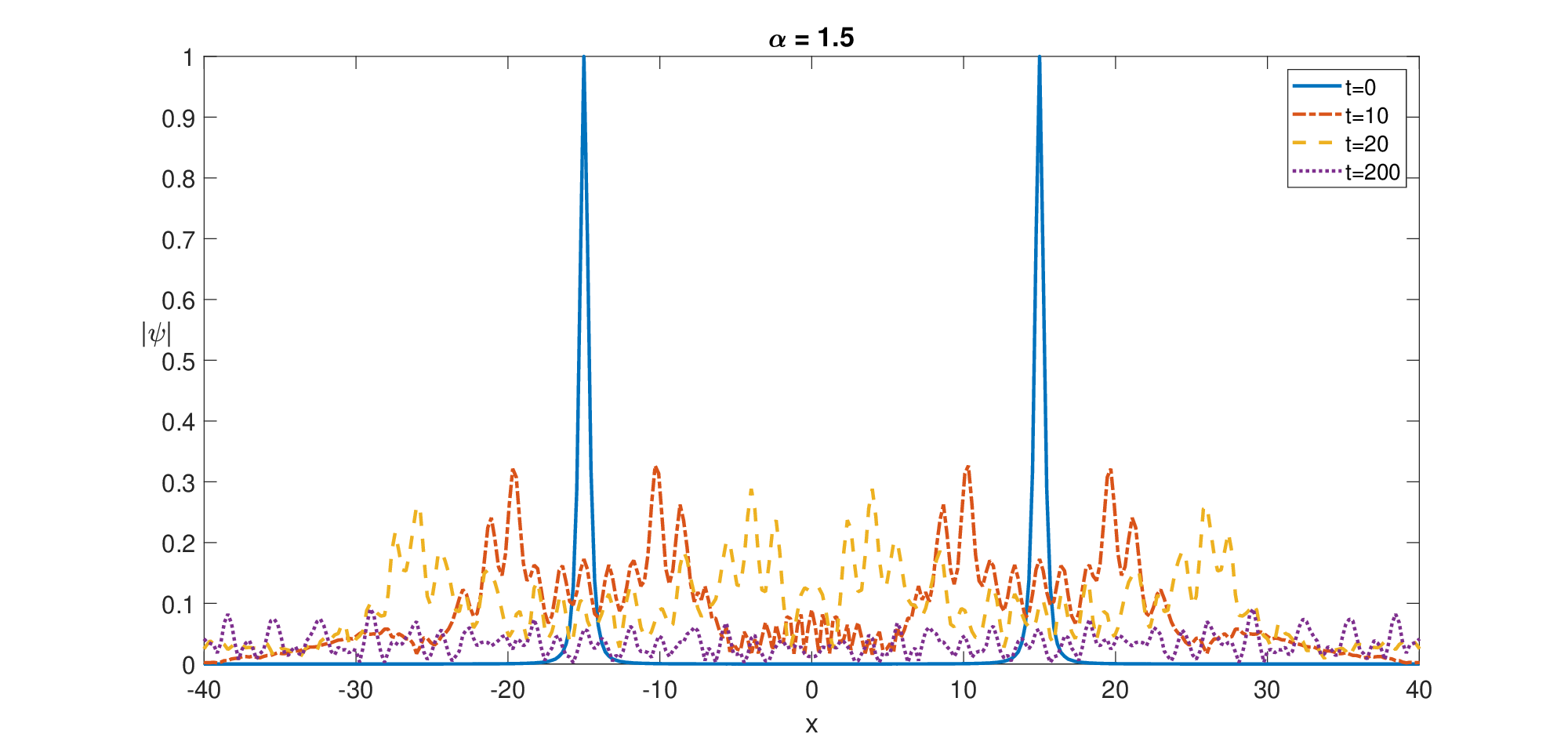}
	\end{center}
	\caption{\small Self-localized two soliton profiles at different times for $\alpha=1.5$ for a photorefractive potential of $V=I_o\cos^2(k_x x)=2\cos^2(2x)$.}
	\label{fig11}
\end{figure}
The two soliton solutions of the fNLSE under the effect of a photorefractive potential in the form of $V=I_o\cos^2(k_x x)=2\cos^2(2x)$ is depicted in Fig.~\ref{fig11}. After comparison of Figs.~\ref{fig6} and \ref{fig11}, it is possible to state that on the shorter time scales short-wavelength undulations on the main soliton profiles are significantly trapped near two initial peaks. The $\mu-P$ graphs and  $t-P$ graphs obtained for the soliton solutions of the fNLSE under the effect of a photorefractive potential are similar and for the sake of brevity are not presented, but the steady-in-the-mean state is achieved earlier around the dimensionless time of $t \approx 100$. They are also stable and robust against noisy perturbations.

\subsection{Results for q-deformed Rosen-Morse Potential, $V=A\text{sech}_q^2(\beta x)+ B\text{tanh}_q(\beta x)$}
Another type of potential function commonly used to effects of shifts and asymmetry in the medium under investigation is the q-deformed potentials~\cite{falaye2013exact,eleuch2018some}. 
Various analytical solutions of the nonlinear systems with q-deformed potential are present in the literature~\cite{falaye2013exact,eleuch2018some}, however the majority of the solutions are valid for models other than fNLSE and valid for steady-state conditions. In order to investigate the effects of q-deformation on the solitons of the fNLSE, we use a q-deformed Rosen-Morse potential. For this purpose, we use a potential function in the form of $V(x)=A\text{sech}_q^2(\beta x)+ B\text{tanh}_q(\beta x)$ where q-deformed hyperbolic functions are formulated as $\text{sech}_q(x)=\frac{2}{e^{x} + q e^{-x}}$ and $\text{tanh}_q(x) =\frac{\text{sinh}_q(x)}{\text{cosh}_q(x)}= \frac{e^{x} - q e^{-x}}{e^{x} + q e^{-x}}$~\cite{falaye2013exact}. 
The parameters of the $q$-deformed potential functions is selected as $A=B=-1, \beta=1, 0.1, q=0.6$ as the typical parameters used in the literature \cite{bayindir2021self,falaye2013exact}.

\begin{figure}[t!]
	\begin{center}
		\hspace*{-0.1\columnwidth}
		\includegraphics[width=1.2\columnwidth]{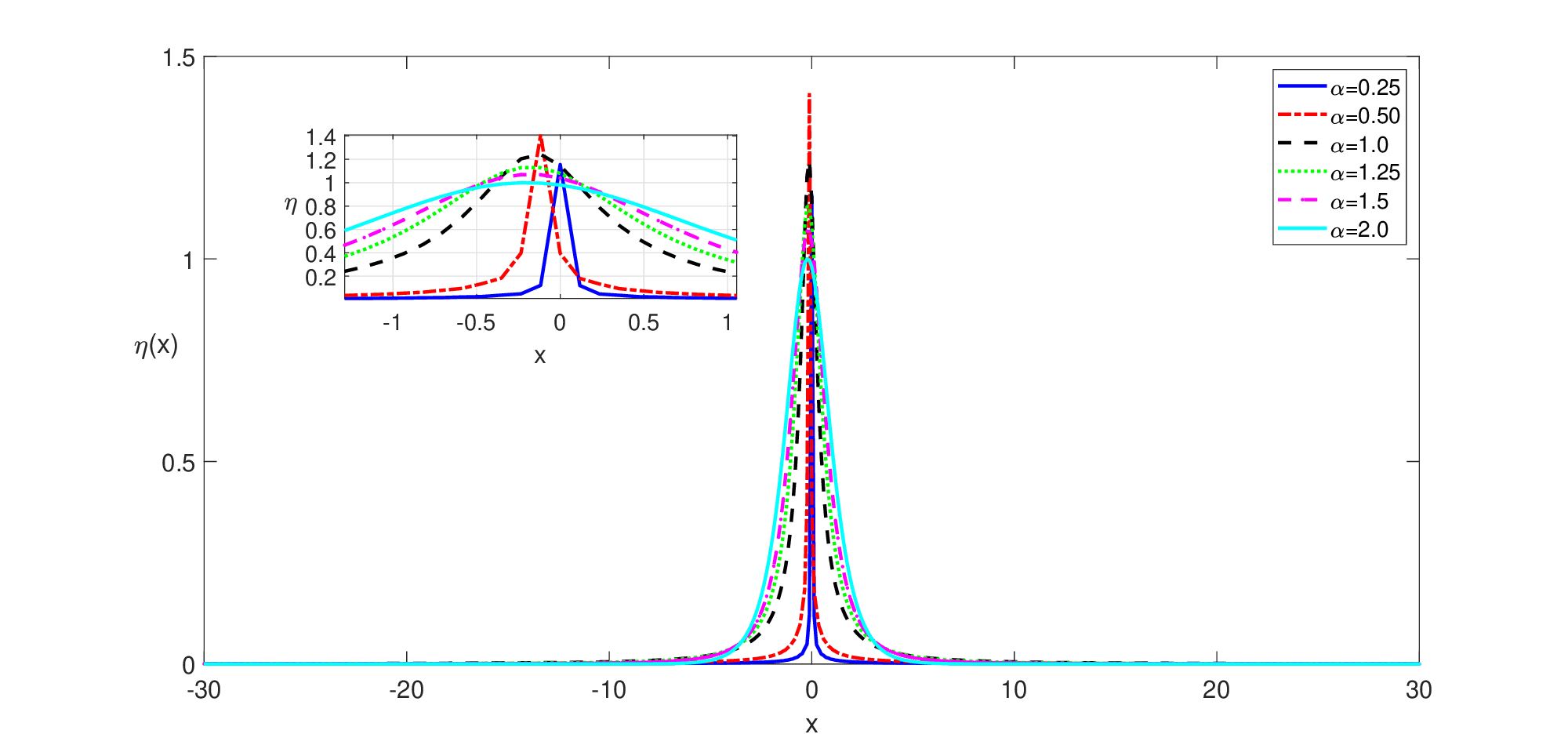}
	\end{center}
	\caption{\small Self-localized one soliton solutions of the fNLSE with a Rosen-Morse potential of $V(x)=A\text{sech}_q^2(\beta x)+ B\text{tanh}_q(\beta x)$ with $q=0.6, \beta=1$ for various values of $\alpha$.}
	\label{fig12}
\end{figure}

In Fig.~\ref{fig12}, we depict the one soliton solutions of the fNLSE with the q-deformed Rosen-Morse potential term with the parameter $\beta=1$ obtained for various values of the fractional-order $\alpha$. Compared to the solitons depicted in Figs.~\ref{fig1} and \ref{fig9} which are obtained for the zero potential and for the photorefractive potential cases respectively, the solitons depicted in Fig.~\ref{fig12} has distinctive features. The dip in the peak or two peaked solitons observed for the photorefractive potentials are not observed for the q-deformed Rosen-Morse and their shapes are more similar to the ones obtained for the zero potential case. However, as an important difference, the location of the peak is shifted and the profile exhibits skewness due to the q-deformed Rosen-Morse type potential used in their construction.

\begin{figure}[t!]
	\begin{center}
		\hspace*{-0.1\columnwidth}
		\includegraphics[width=1.2\columnwidth]{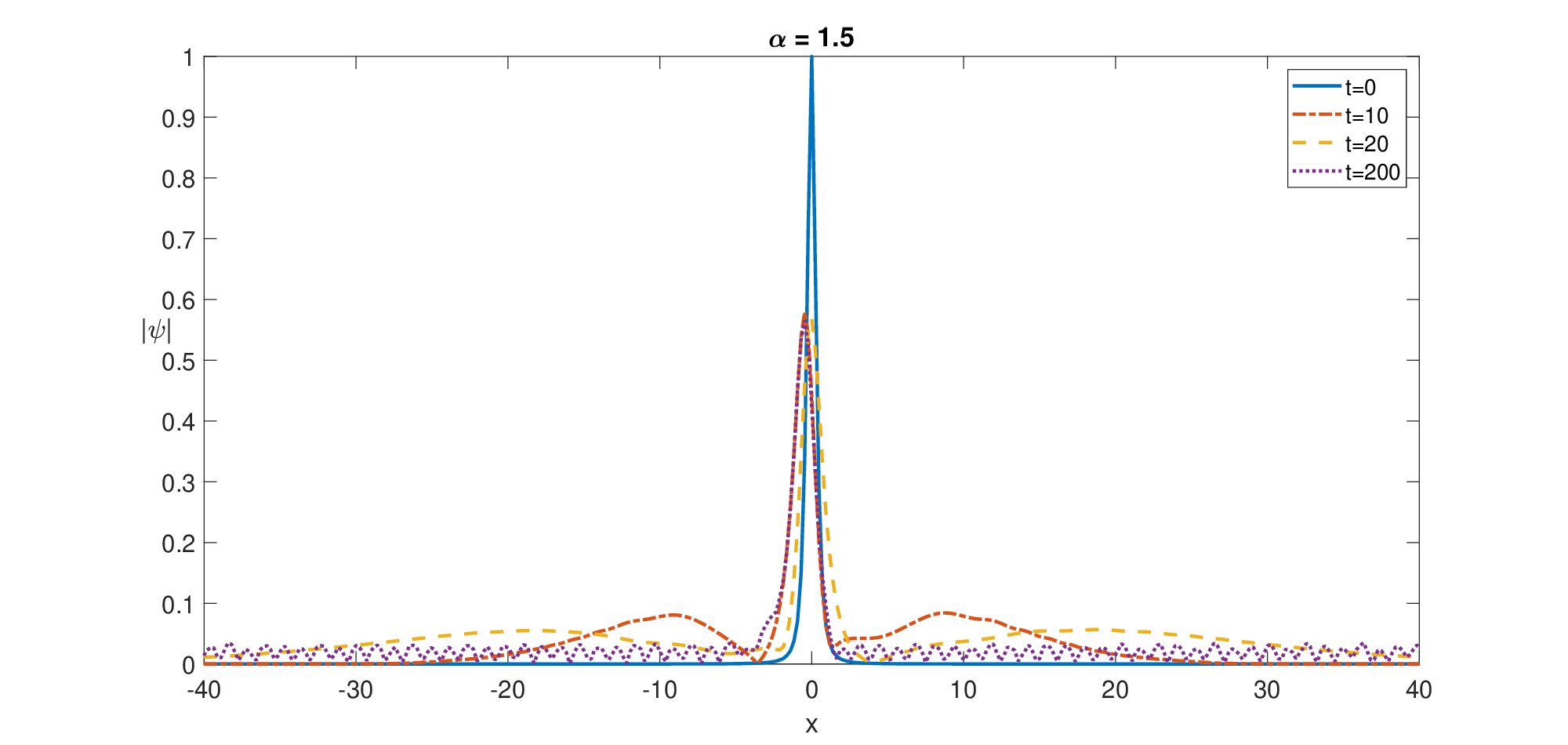}
	\end{center}
	\caption{\small Self-localized one soliton profiles at different times for $\alpha=1.5$ for a Rosen-Morse potential of $V(x)=A\text{sech}_q^2(\beta x)+ B\text{tanh}_q(\beta x)$ with $q=0.6, \beta=1$.}
	\label{fig13}
\end{figure}

After a careful inspection of Fig.~\ref{fig13}, it is possible to state that during the temporal dynamics of the fNLSE with q-deformed Rosen-Morse potential, the shifts in soliton peak positions and their profile skewnesses can be observed. Additionally, about an axis passing through the origin, the wavefield exhibits an asymmetric distribution of the wavefunction. This asymmetry can be useful for the implementation of filtering-tunneling potentials of the optical and quantum processes modeled in the frame of the fNLSE.

\begin{figure}[t!]
	\begin{center}
		\hspace*{-0.1\columnwidth}
		\includegraphics[width=1.2\columnwidth]{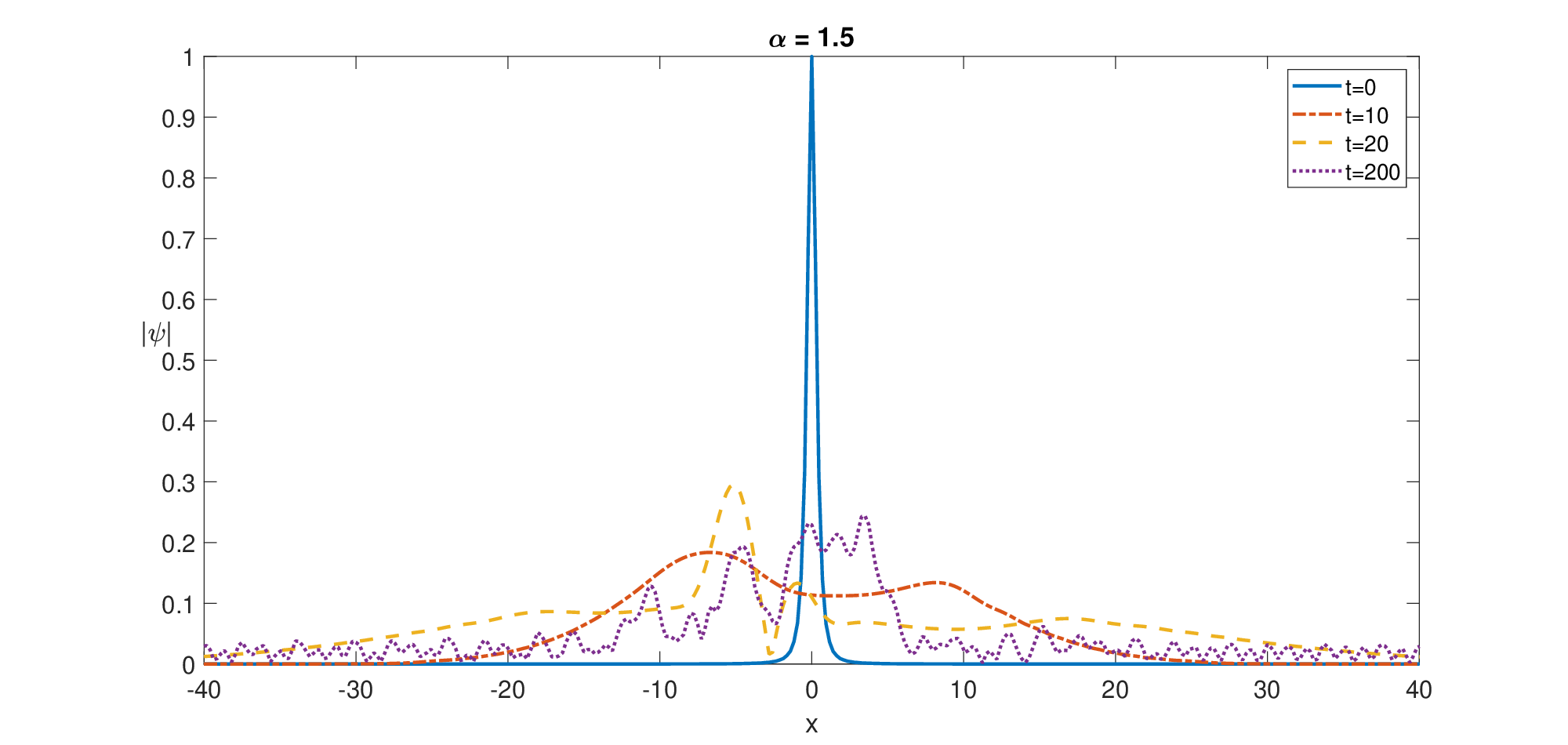}
	\end{center}
	\caption{\small Self-localized one soliton profiles at different times for $\alpha=1.5$ for a Rosen-Morse potential of $V(x)=A\text{sech}_q^2(\beta x)+ B\text{tanh}_q(\beta x)$ with $q=0.6, \beta=0.1$.}
	\label{fig14}
\end{figure}

Lastly, we examine the effect of parameter $\beta$ of the q-deformed Rosen-Morse potential. The value of $\beta$ is reduced from 1 to 0.1, and solitons of the fNLSE obtained by PM and then time-evolved by fractional spectral method with $4^{th}$ order Runge-Kutta time-integrator are depicted in Fig.~\ref{fig14}. Such a reduction in the  $\beta$ parameters leads to a significant increase in the q-deformed Rosen Morse potential, and similar to the photorefractive potential case, low amplitude undulations with shorter wavenumber components become apparent on the main soliton profile. However, skewness and asymmetry of the wavefield are still observed.

\section{Conclusion}
In this paper, we extended the Petviashvili method (PM) which can be used to construct the soliton solutions of the fractional nonlinear Schr\"{o}dinger equation (fNLSE). We investigated the characteristics and the effects of fractional order processes on those solitons. Additionally, the stabilities and temporal dynamics of those solitons of the fNLSE are investigated via a Fourier spectral algorithm with a $4^{th}$ order Runge-Kutta time integrator, where the fractional derivatives are computed spectrally. We showed that the fractional derivative order parameter $\alpha$ controls the slenderness of the soliton peak profiles, soliton profiles more slender as $\alpha$ reduced. We showed that under zero potential, photorefractive potential, and q-deformed Rosen-Morse potential, the fNLSE admits soliton solutions. These solitons exhibit a splitting and spreading behavior, however, various forms of the potentials can be used to alter their dynamics. We also showed that those solitons are robust against noisy perturbations and the effect of noise on these solitons can be removed by a Savitzky-Golay filter. For the zero potential case, solitons of the fNLSE split into two pieces initially, then spread. However, in the photorefractive and q-deformed Rosen-Morse potential cases shorter wavelength components are generated. This effect is more prominent in the case of the photorefractive potential. For the q-deformed Rosen-Morse potential case, the shifts in soliton peak location and skewness in their profiles are observed.

The fractional-order soliton dynamics reported in this work are promising. The fractional-order derivative controls the slenderness of the solitons and their peakedness. Thus, the physical processes described by fractional derivatives can be used to fine-tune the soliton profiles to increase the resolution of the solutions, enabling pulse pinpointing and matched filtering with a better performance. Also, the photorefractive and q-deformed potentials can be used for multi-frequency wavefield generation as well as tunneling, shifting, and adjusting the skewness of those solitons. Our findings can be used in many disciplines, including but are not limited to quantum physics, nonlinear optics and hydrodynamics, Bose-Einstein condensation, and quantum control theory. 

There are many possible research directions based on the research presented in this paper. First of all, our results can be easily generalized to N-soliton solutions. The effects of other types of potentials of fNLSE can also be investigated within the frame of the approach presented herein. Some of these potentials are Eckart, Coulomb, Yukawa, parity-time symmetric Rosen-Morse, $\epsilon$-deformed an-harmonic potentials. Fractional order equations with other forms of nonlinearities (i.e.,~\cite{bayindir2016rogueChaotic,bayindir2016rogueSpectra}) can also be studied within our numerical framework. Such future works on this subject would contribute to significant improvements and new insights in various branches of science, some of which are mentioned above.



\section*{Declarations}

\begin{itemize}
	\item \textbf{Funding} F.O. acknowledges the Personal Research Fund of Tokyo International University.
	\item \textbf{Conflict of interest/Competing interests} We have no competing interests.
	\item \textbf{Ethics approval}  Not applicable 
	\item \textbf{Consent to participate}	 Not applicable
	\item \textbf{Consent for publication}	 Not applicable
	\item \textbf{Availability of data and materials}	 Not applicable
	\item \textbf{Code availability}	 Not applicable 
	\item \textbf{Authors' contributions}	 Not applicable
\end{itemize}

\bigskip
\begin{flushleft}%
\end{flushleft}


\bibliography{OQuLv220923}


\end{document}